\documentstyle[prl,aps,multicol, epsfig]{revtex}

\begin{document}
\draft
\widetext

\title{Anomalous Magnetic Field Dependence of the Thermodynamic Transition Line in 
the Isotropic Superconductor (K,Ba)BiO$_3$}

\author{S.~Blanchard$^1$, T.~Klein$^1$, J.~Marcus$^1$, I.~Joumard$^1$, A.~Sulpice$^2$,
 P.~Szabo$^3$, P.~Samuely$^3$, A.G.M.~Jansen$^4$ and C.~Marcenat$^5$}
\address{$1$- Laboratoire d'Etudes des Propri\'et\'es Electroniques des
Solides, Centre National de la Recherche Scientifique, BP166, 38042 Grenoble 
Cedex 9, France}
\address{$2$- Centre de Recherche sur les Tr\`es Basses Temp\' eratures,
 Centre National de la Recherche Scientifique, BP166, 38042 Grenoble 
Cedex 9, France}
\address{$3$- Institute of Experimental  Physics, Slovak Academy of
Sciences, SK-04353 Ko\v{s}ice, Slovakia}
\address{$4$- Grenoble    High     Magnetic    Field    Laboratory,
Max-Planck-Institut   f\"{u}r   Festk\"{o}rperforschung   and   
Centre National de  la Recherche Scientifique,  B.P. 166, F-38042  Grenoble
Cedex 9, France}
\address{$5$- Commissariat \`a l'Energie Atomique - Grenoble, 
D\'epartement de Recherche Fondamentale sur la Mati\`ere Condens\'ee, 
SPSMS, 17 rue des Martyrs, 38054 Grenoble Cedex 9, France}

\date{today}
\maketitle

\widetext
\begin{abstract}
Thermodynamic (specific heat, reversible magnetization, tunneling
spectroscopy) and transport measurements have been performed on high quality
(K,Ba)BiO$_3$ single crystals. The temperature dependence of the magnetic
field $H_{Cp}$ corresponding to the onset of the specific heat anomaly presents
a clear positive curvature. $H_{Cp}$ is significantly smaller than
the field $H_\Delta$ for which the superconducting gap vanishes but is closely related to 
the irreversibility line deduced from transport data. Moreover, the temperature dependence of the reversible
magnetization present a strong deviation from the Ginzburg--Landau theory
emphasazing the peculiar nature of the superconducting transition in this
material.
\end{abstract}

\pacs{PACS numbers: 7425Bt, 7425Dw, 7460Ec}

\begin{multicols}{2}

\narrowtext

The $H-T$ phase diagram of high $T_c$ superconducting oxides has
been the focus of intense theoretical and experimental study
during the past decade \cite{Blatter94}. One of the most
interesting phenomemon which has been observed is the existence of
a melting line $T_m(H)$ above which the vortex lattice melts into
a liquid of entangled lines \cite{Blatter94}. This melting has
very important consequences for the physics of vortices, since the
free motion of the flux lines in the liquid state gives rise to a
large dissipation and renders the system useless for applications.
The presence of this liquid phase also hinders any direct
determination of the upper critical field $H_{c2}$ from standard
magnetotransport data. Indeed, any resistive "critical" field
$H_R(T)$ defined as the field for which the resistivity reaches
some arbitrary value $R$ (e.g. $50\%$ of the normal state
resistivity), then presents a positive curvature in high $T_c$ cuprates
\cite{Transpcuprates} as well as in (K,Ba)BiO$_3$
\cite{Samuely98}, \cite{Klein99}. \

Similarly, the presence of strong thermal fluctuations also complicates the
determination of $H_{c2}$ from thermodynamic measurements.
These fluctuations lead to broad and smooth anomalies in specific heat
measurements ($C_p$). Thus the field $H_{Cp}$ (defined for instance
by the inflexion
point of $C_p/T$) can present either a positive curvature in
Y$_1$Ba$_2$Cu$_3$O$_{7 - \delta}$ (YBCO) crystals
\cite{Roulin98}, a linear dependence in Tl$_2$Ba$_2$Cu0$_6$
\cite{Carrington96} or
even almost no dependence with field in highly anisotropic systems
such as Bi$_2$Sr$_2$CaCu$_2$O$_8$ \cite{Junod99} and
HgBa$_2$Ca$_2$Cu$_3$O$_{8 + \delta}$ \cite{Marcenat96}.
It is still unclear whether this upper critical field still exists as
a transition line or is just some smooth crossover between the vortex
liquid and the normal state. In order to shed light on this
issue, we performed thermodynamic (specific heat, magneto-tunneling,
reversible magnetization)
and transport measurements on high quality optimally doped $(K_x,Ba_{1-x})BiO_3$ 
single crystals (x $\sim$ 0.4). \

\begin{figure}[tbp]
    \centerline{
    \epsfxsize 7cm
    \epsffile{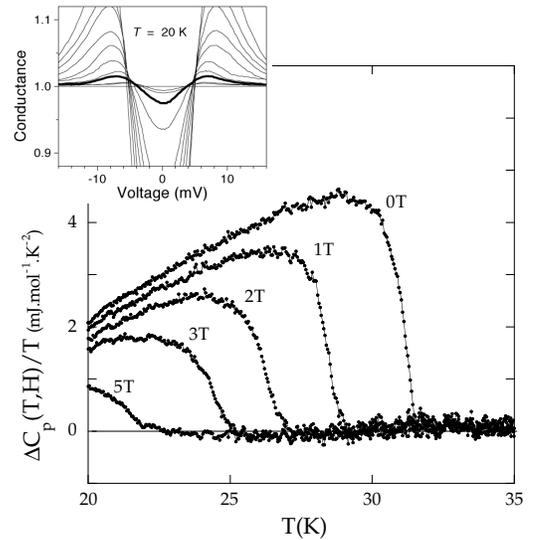}}
    \caption{Temperature dependence of the specific heat anomaly in a
(K,Ba)BiO$_3$ single crystal. $\Delta$ $C_p$ = $C_p$(H,T) - 
$C_p$(H=7T,T). Inset: magnetic field dependence of the tunneling 
spectra at $T = 20$K and $H(T) = 0, 1, 2, 3, 4, 5, 6, 7$ 
(bold curve), $8, 9, 10$}
\end{figure}

Thermal fluctuations are small in this system (the Ginzburg number
$G_i$ is only $\sim 10^{-4}$) due to its isotropic structure,
a $T_c \sim 31$K, and a coherence length on the order of $30$ \AA.
Nevertheless, some of us have previously shown that the
magneto-transport data is well described by the vortex-glass scaling formalism
\cite{Klein99} suggesting that the
vortex solid melts into a liquid at high temperature through a second
order phase transition. In the following, the vortex-glass transition
field $H_{vg}$ is
defined as the field for which $R \rightarrow 0$. On the other hand,
magnetotunneling measurements have shown that the superconducting
gap $\Delta$
closes above some characteristic field $H_\Delta$. The corresponding
$H_\Delta(T)$ line is in reasonable agreement with the classical
Werthamer-Helfang-Hohenberg (WHH) theory for the upper critical field
\cite{Samuely98}. The
physical image emerging from those measurements was then that of a liquid phase
existing for $H_{vg} < H < H_\Delta \sim H_{c2}$. However, we will
show here that
the magnetic field $H_{Cp}$ is {\it smaller} than $H_\Delta$ and
presents a strong {\it positive curvature}, 
indicating a very peculiar nature for the thermodynamic superconducting transition.\

The measurements were performed on new particularly homogeneous
single crystals presenting very sharp superconducting transitions
in both transport ($\Delta T_c \sim 0.15$K) and ac susceptibility
($\Delta T_c \sim 0.2$K for $h_{ac} < 0.01$G) measurements. The
specific heat was measured by an ac technique \cite{Sullivan68}
which allows us to measure small samples (here a few $10^{-2}$
mm$^3$) with high sensitivity (typically $1$ part in $10^4$). Heat
was supplied to the sample at a frequency $\omega$ on the order of
a few Hz by a light emitting diode via an optical fiber. The
induced temperature oscillations were measured by a
chromel-constantan thermocouple, which was calibrated in situ
using a very pure silver single crystal as a reference.  These
measurements are only relative and were renormalized using the
data from ref \cite{Woodfield99}. Figure 1 displays the
temperature dependence of the specific heat at various magnetic
fields up to $5$T. The $7$T curve is equal to the normal state
specific heat above $T = T_c(H=7T) \sim 20$K and has been used as
a base line in Fig.1.\

The unusally high quality of our crystals is attested to by the narrow width of the transition,
which is on the order of $1K$ in zero magnetic field, and by  the amplitude of the specific
heat jump, $\Delta C_p(T_c)/T_c \sim 4-5$ mJ.mol$^{-1}$K$^{-2}$.
This ratio is $\sim$ 2 and 5 times larger than that of refs.
\cite{Woodfield99}and \cite{Graebner89}, respectively, which are the only prior reports of a
specific heat anomaly at $T_c$ in this system. This allowed us to
study  the magnetic field dependence of the specific heat anomaly in much greater detail.
As shown in Fig.1, this anomaly remains well defined in magnetic
fields. The $T_{Cp}(H)$ curve corresponding to any characteristic
point of the transition (e.g. the onset, mid-point or maximum)
presents a clear positive curvature. This can be seen in Fig.2
where $T_{Cp}$ corresponds to the onset of the peak. The shaded
area represents the temperature difference between the onset and
the mid-point of the transition. A very similar upward curvature
has been obtained in two other samples. A positive curvature was 
already obtained by transport \cite{Klein99} and magnetic measurements 
\cite{Magnetic}. However, 
in non-classical superconductors, there is an ambiguity in defining $H_{c2}$ 
using those techniques. \

Another striking feature of the specific heat anomaly is the rapid
collapse of its height with field. Such a behaviour was
previously observed in high $T_c$ materials \cite{Junod99}
\cite{Marcenat96} and is
usually attributed to the presence of highly field dependent thermal
fluctuations. The observation of a similar collapse in $(K,Ba)BiO_3$
remains a puzzling issue. Geometrical arguments based on
entropy conservation imply that this reduction must be accompanied by
a rapid, anomalous, increase of the specific heat at low temperature
for fields much smaller
than the upper critical field of $\sim 30$T deduced from transport
data. The data from ref \cite{Woodfield99} also suggest such a behaviour.\

\begin{figure}[tbp]
    \centerline{
    \epsfxsize 7cm
    \epsffile{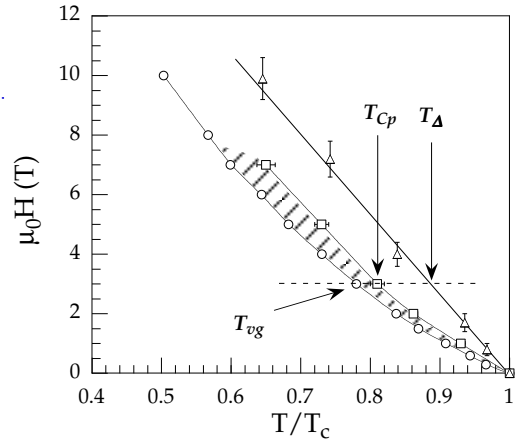}}
    \caption{$H-T$ phase diagram of the (K,Ba)BiO$_3$ system.
Squares ($H_{Cp}(T)$): onset
on the specific heat anomaly (see Fig.1, the shaded area
represents the temperature difference between the onset and the
mid-point of the
anomaly), circles ($H_{vg}(T)$ : "vortex-glass transition" line
deduced from transport
measurements ($R \rightarrow 0$), triangles: $H_\Delta (T)$ defined as
the line for which the superconducting gap measures by tunneling
spectroscopy vanishes. The different characteristic temperatures at $3$T
(arrows) as shown in Fig.3.}
\end{figure}

As pointed out above, the specific heat measurements are in striking contrast
with our previous magneto-tunneling \cite{Samuely98} data which
suggested than the upper critical field
(defined as the field $H_\Delta$ for which the superconducting gap is
completely closed) presents a classical WHH dependence. We have thus 
performed similar magneto-tunneling measurements on the sample which has 
been used in the specific heat experiments. The inset of Fig.1 shows the 
evolution of the tunneling spectra near the normal state at $T = 20$K with increasing 
magnetic fields. $H_\Delta$ has been defined as the field for which the 
superconducting gap is compeltely closed ($H_\Delta(20K) \sim 10$T). As 
shown by the bold curve, at $H=H_{Cp}(20K) \sim 7$T, a gap some feature remains clearly visible in 
the corrersponding tunneling spectrum (bold curve). As showis still well 
developed and it closes at about $10$T. As shown in Fig.2, in agreement with
our previous data, the curvature is much less
pronounced for $H_\Delta$ and surprisingly
$H_{vg} < H_{Cp} < H_\Delta$. Note than even though the difference between
$T_{Cp}$ and $T_\Delta$ is quite large, the resistivity reaches about $95\%$ of
its normal state value ($R_N$) at $T_{Cp}$ and smoothly increases up
to $\sim 100 \%$ for $T \sim T_\Delta$ (see Fig.3).
The specific heat anomaly thus defines a fundamental boundary below which the
resistivity drops rapidly towards zero but superconductivity is only completely
destroyed at $T = T_\Delta$ for which $R = R_N$ and the superconducting gap is
completely suppressed. Note that the difference between $T_\Delta$ and
$T_{Cp}$ is {\it not} due to sample homogeneities since heavy
ion irradiation leads to an increase of $T_{Cp}$ which then tends towards
$T_\Delta$ \cite{Irradiation}. This difference could be due to the 
existence of a pseudo-gap but, in contrast to cuprates, this pseudo-gap 
would only occur in non zero magnetic field.\

\begin{figure}[tbp]
    \centerline{
    \epsfxsize 7cm
    \epsffile{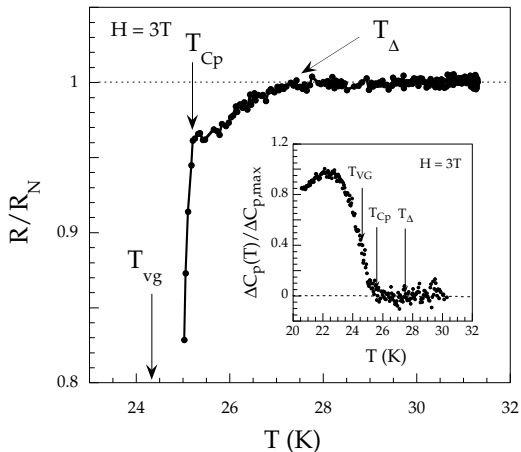}}
    \caption{Resistive transition at $H = 3$T. The onset of the
specific heat
anomaly (see Fig.1) corresponds to the temperature below which the
resistivity rapidly decreases. $R \rightarrow R_N$ for $T = T_\Delta$
(temperature for which the
superconducting gap closes, $R_N$ is the
normal state resistivity). In the insert: renormalized specific heat
vs temperature. As shown $T_{vg}$ is close to the mid-point of the
specific heat anomaly.}
\end{figure}

To complete this study of the thermodynamic properties, we have performed
reversible magnetization measurements ($M_{rev}$) using a SQUID
magnetometer. In the intermediate magnetic field range,
$H_{c1} \ll H \ll H_{c2}$, the reversible magnetization of extreme type-II
superconductors in the London model \cite{London88} is given by : $
M_{rev}~=~-\frac{\alpha \Phi_0}{32 \pi \lambda^2} \ln \frac{\beta
H_{c2}(T)}{H}$
where $\alpha \sim 1$, $\Phi_0$ is the flux quantum, $\lambda(T)$ the
magnetic penetration depth, $H_{c2}(T)$ the upper critical field and
$\beta \sim 0.37$ \cite{Hao91}. Thus, $\partial M_{rev}/\partial $ln$(H)$ is expected to be 
proportional to $1/\lambda^2(T) \sim (1-T/T_c)$ close to $T_c$.
$M_{rev}$ presents a clear logarithmic dependence (see \cite{Joumard00})
and the corresponding $\partial M_{rev}/\partial $ln$(H)$ are shown
in Fig.4. However, as shown, strong deviations from the expected
linear behaviour are visible
and $\partial M_{rev}/\partial $ln$(H)$ can be described  much better
by a $(1- T/T_c)^{1.5}$
law (solid line, the dotted line is a WHH dependence for $1/\lambda^2$).
A similar deviation  is usually observed in
high $T_c$ cuprates \cite{MrevCuprates} and has been attributed to the
presence of strong fluctuations. It was originally suggested
by Nelson {\it et al.} \cite{Nelson89} that elastic distortions of the
vortex lattice (in the presence of strong thermal fluctuations) may lead to a
significant contribution to the entropy of the
system. However the corresponding contribution to $\partial
M_{rev}/\partial $ln$(H)$ is
expected to be of the order of $k_B T/\Phi_0 \xi$
\cite{Bulaievski92}(where $k_B$ is the Boltzmann constant),
  which is about one order of magnitude smaller than the observed
deviation in $(K,Ba)BiO_3$ \cite{Crossingpoint}.\

Both specific heat and reversible magnetization data thus point out
the peculiar nature of the superconducting transition in (K,Ba)BiO$_3$.
Moreover, as shown in Fig.2, $H_{vg}$ and $H_{Cp}$ present
a very similar temperature dependence emphasizing the close relation
between those two lines. As shown in the inset of Fig.3, the $R=0$ criterion is close
to the mid-point of the $C_p$ transition.
This then suggests two possibilities: either (i) the positive curvature in
$H_{R}$ is {\it not} due to the melting of the vortex solid as usually
suggested but is an intrinsic property of the $H_{c2}$ line ($=
H_{Cp}$) or (ii) the specific heat anomaly is
not related to $H_{c2}$ but instead marks a transition in the vortex state. \
Many theoretical models have
been developped in order to explain a "possible" upward curvature of the upper
critical field in cuprates (including a spin-charge separation \cite{Dias94}
  or very strong coupling \cite{Marsiglio87}) but none of those models can
be applied to the (K,Ba)BiO$_3$ system. A positive curvature has also been
predicted by Ovchinnikov and Kresin \cite{Ovchinnikov94} in the
presence of inhomogeneities and/or
magnetic impurities. However, in this case this curvature is only expected to
appear at low temperature (i.e. when the superconducting coherence length
becomes smaller or on the order of the size of the impurity) and
$H_{c2}$ is still expected to vary linearly with $T$ close to $T_c$. 
Finally, the condensation of charged bosons in magnetic field is also 
expected to lead to a positive curvature \cite{Alexandrov93}.  But, one 
has then to take into account the fact that we do not observe any feature 
in the tunneling spectra above $T_{Cp}$ in zero magnetic field.  
On the other hand, a second scenario is that $H_{c2} \sim H_\Delta$
and that the specific heat anomaly marks a transition
within the vortex state. For instance, specific heat anomalies, both peaks and/or steps, have been
observed for transitions in the vortex state in YBCO \cite{Melting}
\cite{Bouquet01}. However, even though the amplitude of $\Delta C_p$ per unit volume measured here
is similar to the one associated with vortex  melting in YBCO, its shape 
 is quite different and looks much like a classical superconducting transition.
 
In summary, specific heat and reversible magnetization measurements
demonstrate the very peculiar nature of the thermodynamic
superconducting transition in
the cubic (K,Ba)BiO$_3$ system. The strong deviations from the Ginzburg--Landau model in
the reversible magnetization, the rapid collapse of the specific heat
anomaly with magnetic field and the positive curvature in $H_{vg}(T)$ associated with a
vortex-glass scaling behaviour, all suggest the presence of {\it strong}
fluctuations. It is interesting to note that the
superconducting transition in (K,Ba)BiO$_3$ occurs in the vicinity of
a metal - insulator
transition and the carrier density is very small in this system. This might
lead to the presence of large {\it quantum} fluctuations
\cite{Ivlev94} \cite{Emery95}. The field $H_\Delta$ for which the
superconducting gap completely disappears in tunneling spectroscopy is
larger than the field $H_{Cp}$ corresponding to the onset of the specific
heat anomaly. $H_{Cp}$ (as well as any characteristic field deduced from
transport measurements) presents a clear positive curvature.\

\begin{figure}[tbp]
    \centerline{
    \epsfxsize 7cm
    \epsffile{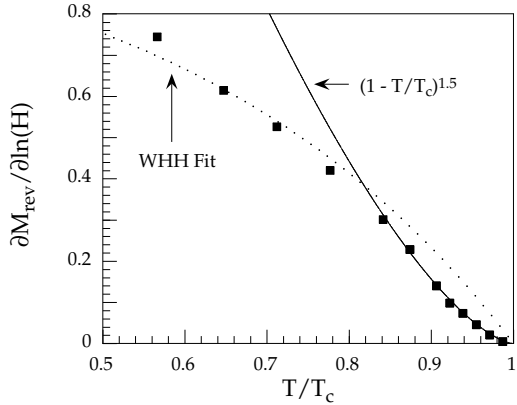}}
    \caption{Temperature dependence of $\partial M_{rev}/\partial
$ln$(H)$ in a
(K,Ba)BiO$_3$ single crystal.
The dotted line is a WHH fit for the temperature dependence of
$1/\lambda^2$ (see text for details). The solid line is a $(1 - T/T_c)^{1.5}$
fit to the data.}
\end{figure}

We would like to thank L.Paulius for interesting discusions and
N.E.Phillips and R.A.Fischer for sending us their specific heat data on
(K,Ba)BiO$_3$.

\end{multicols}

\begin{references}
\bibitem{Crossingpoint} The main difference between (K,Ba)BiO$_3$ and cuprates
is that $\partial M_{rev}/\partial $ln$(H)$ remains positive on the entire
temperature range in (K,Ba)BiO$_3$.
\end{references}
\end{document}